\documentclass{elsart}
\journal{Eur. Phys. J. Appl. Phys.}

\usepackage{epsfig}
\usepackage{amsmath}
\newcommand{\degree}   {$^{\circ}$}

\begin{document}
\begin{frontmatter}
\title{Measurement of the temperature dependence of pulse lengths in an n-type germanium detector}

\author[a]{I.~Abt\corauthref{cor}\ead{isa@mppmu.mpg.de}},
\author[a]{A.~Caldwell},
\author[a,b]{J.~Liu},
\author[a]{B.~Majorovits},
\author[a]{O.~Volynets}
\address[a]{Max-Planck-Institut f\"ur Physik, M\"unchen, Germany}
\address[b]{now: Institute for the Physics and Mathematics of the Universe, Tokyo University, Tokyo, Japan}

\corauth[cor]{Max-Planck-Institut f\"ur Physik, F\"ohringer Ring 6, 
              80805 M\"unchen, Germany, 
              Tel.: +49-(0)89-32354-295, FAX: +49-(0)89-32354-528}

\date{\today}
\begin{abstract}
The temperature dependence of the pulse length 
was measured 
for an 18--fold segmented  n-type germanium detector
in the temperature range of 77 -- 120\,K. 
The interactions of 122\,keV photons originating from
a $^{152}$Eu source were selected and pulses as observed on the core
and segment electrodes were studied.
In both cases, the temperature dependence 
can be well described by a Boltzmann-like $ansatz$.

\end{abstract}
\end{frontmatter}
\maketitle
%

\section{Introduction}
\label{section:introduction}

High purity germanium detectors, HPGeDs, 
are widely used in spectroscopy, for example in
the detection of low levels of radioactivity.
They are also  used in a wide variety 
of applications in particle and nuclear physics \cite{Kno99,eberth}. 
In particular, they are used for gamma ray tracking 
in arrays like AGATA \cite{agata+greta} and GRETA \cite{agata+greta}, 
and in the
search for neutrinoless double beta decay in experiments like
GERDA \cite{gerda} and MAJORANA \cite{majorana}.

For some applications, the shapes of the electric pulses collected
on the electrodes are of interest. 
They are used in so called
pulse shape analyses which are usually used to obtain information
about event topologies. This can either help in gamma ray tracking \cite{psag}
or in identifying background events \cite{psam,psa}. 

A fundamental parameter
of a pulse is its length. 
It can be used to measure the distance of the energy deposit
from the electrodes.
Its absolute value, however, depends also on the impurity density
of the crystal and the relative location of the energy deposit with
respect to the crystal axes. Other important factors are the
mobilities, $\mu$, of the charge carriers,
electrons and holes, which are temperature, $T$, dependent.
 
The mobilities of the charge carriers have been 
measured in the past. However, the results,
obtained for different samples, do not completely agree \cite{bart}.
In principle, the mobilities can be determined by comparing
measured to simulated pulses. In practice, this is
difficult, because the impurity density distribution
of a detector crystal is also
not precisely known. 

The $T$ dependence of the mobilities, can be directly related
to the $T$ dependence of the pulse lengths.
Photon interactions from
the 122\,keV line of a $^{152}$Eu source were used to select charge 
depositions close to
the outer surface of an n-type cylindrical true-coaxial  germanium detector. 
Holes created close to the surface quickly 
drift to the mantle electrode  and
the electrons drifting to the core 
determine the total length of the pulse. 
The holes influence only the
first part of the pulse. 
Pulses were studied  separately for interactions
close to the crystallographic axes $\langle 110 \rangle$ 
and $\langle 100 \rangle$.

\section{Experimental Setup}
\label{s:exp}

The detector used was a cylindrical true-coaxial 
18--fold (6$\phi$, 3$z$) segmented n-type detector
with a height of 70\,mm and a diameter of 75\,mm. 
It was manufactured by Canberra France and is similar to
the detector previously characterized in detail \cite{char}.
The inner bore had
a diameter of 10\,mm. The density of electrically active impurities  was
given  by the manufacturer as
$0.35\cdot10^{10}/ \mbox{cm}^3$ at the top 
and $0.55\cdot 10^{10}/ \mbox{cm}^3$ at the bottom of the detector.
The change in impurities is assumed to be linear with height, $z$.
The operational voltage was 2000\,V.

The detector was mounted either inside a vacuum cryostat \cite{char}, K1,
or submerged in a liquid nitrogen volume \cite{cold}, GII. 
In both cases, a collimated 
40\,kBq  $^{152}$Eu source with a 
1$\sigma$ beam spot diameter of about 5\,mm 
was used. 
The coordinate system used was cylindrical
with the origin at the geometric center of the detector and 
$\phi=0$ along its crystallographic $\langle 110 \rangle$ axis.
                  
The detector $T$ inside GII  was 
constantly 77.4\,K while it was increasing with time inside K1,
where it was cooled via
a cooling finger submerged in a liquid nitrogen reservoir. When
the reservoir was not refilled, the detector slowly warmed up.
The temperature, $T_{\text{mon}}$, was monitored with a thermal resistor, 
PT100, mounted
as close as possible to the detector inside K1.  
The temperature, $T$, of the detector was
expected to be higher by $\Delta T = T - T_{\text{mon}}$  between  
4 -- 10\,K.

The following datasets were analysed:
\begin{itemize}
  \item DS1: detector operated in GII, submerged in LN2, $T=77.4$\,K;
  \item DS2: detector operated in K1, vacuum, 
             $95$\,K $<T_{\text{mon}}<100$\,K; 
  \item DS3: detector operated in K1, vacuum, 
             $100$\,K $<T_{\text{mon}}<120$\,K. 
\end{itemize}
 
\begin{figure}[htbp]
\centering 
\includegraphics[width=0.8\textwidth]{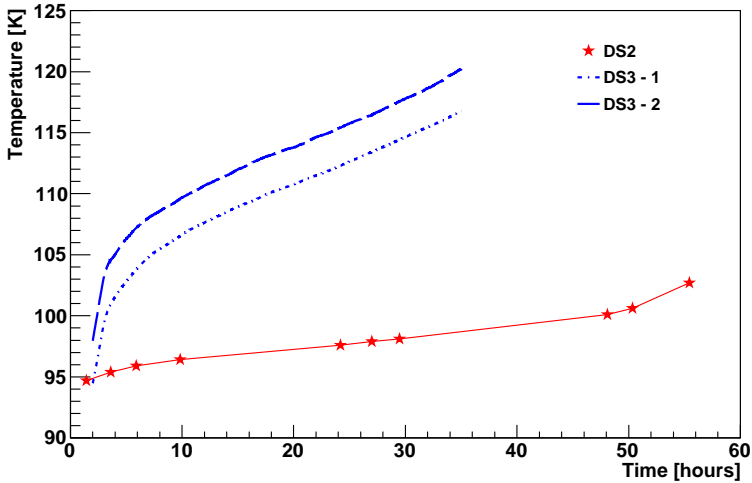}
\caption{Development of $T_{\text{mon}}$ for one of the measurements 
of DS2 and two test warm-ups for DS3.}
\label{f:Tdev} 
\end{figure}

The measurements in K1 were started after the initial cool-down
of the setup.
They were performed while the detector was warming up.
For DS2, $T_{\text{mon}}$ was monitored during breaks 
in the data taking. The value of  $T_{\text{mon}}$
is interpolated for intermediate times.
For DS3, the time dependence of $T_{\text{mon}}$ 
was taken from test warm-ups 
where  $T_{\text{mon}}$ was measured in one minute intervals.
Figure~\ref{f:Tdev} shows the temperature development
for one measurement of DS2 and two test warm-ups
for DS3, one at the beginning and one at the end of the DS3 data taking.

The absolute value of $T_{\text{mon}}$ for DS3 depended on the initial
$T_{\text{mon}}$. The development of $T_{\text{mon}}$ with time
was however stable. 
The $T_{\text{mon}}$ could thus be extrapolated
from the initial $T_{\text{mon}}$ for each measurement cycle.

Data were grouped into 15--minute intervals. 
For the analysis, it was assumed that $T$ and  $T_{\text{mon}}$ 
were constant during these intervals. At the beginning of
the measurements in DS3 the steep rise in $T_{\text{mon}}$ was
excluded.

\begin{figure}[htbp]
\centering 
\includegraphics[width=0.8\textwidth]{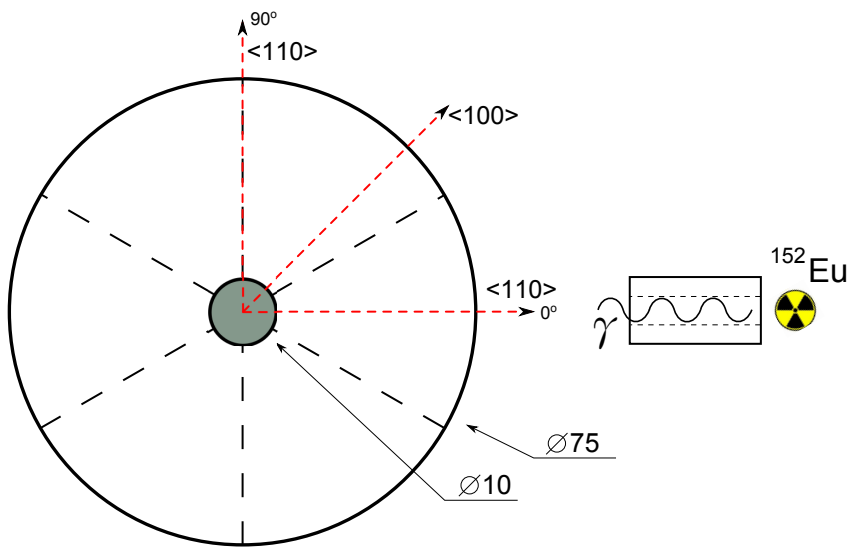}
\caption{Schematic of the measurement setup. Long- and short-dashed lines 
indicate the segment boundaries and crystallographic axes.}
\label{f:setup} 
\end{figure}

Measurements were taken separately around the $\langle 110 \rangle$
axis at $\phi=0^{\circ}$
and the $\langle 100 \rangle$ axis at $\phi=45^{\circ}$.
The measurements for DS3
were $5^{\circ}$ off axis. 
The restrictions of the GII setup did not allow a measurement
on the  $\langle 100 \rangle$ axis. 
Additional data were taken for cross-checks.
Table~\ref{t:DS} provides an overview. 
The two axes were located in different segments of the detector.
None of the axes was close to a segment boundary.
A schematic of the setup is
shown in Fig.~\ref{f:setup}.

\begin{table}[!h]
  \centering
    \begin{tabular}{l | l| r| r| r}
      \hline
      \hline
          & orientation & DS1 & DS2 & DS3\\
      \hline
      \hline
       A1 &$\langle 110 \rangle$   & 0\,\degree & 0\,\degree & 5\,\degree\\
      \hline
       A2 &$\langle 100 \rangle$   &   & 45\,\degree & 50\,\degree\\
      \hline
       \hline
       C1 &$\langle 110 \rangle$  +5\,\degree &   5\,\degree &   &   5\,\degree\\
       \hline
       C2 &$\langle 110 \rangle$ -15\,\degree & -15\,\degree &   & -15\,\degree\\
       \hline
    \end{tabular}
  \vskip 0.2cm
  \caption{Datasets A1 and A2, entering the analysis, and C1 and C2, used for
           cross checks}
  \label{t:DS}
\end{table}

The core and all 18 segments, 19 channels, were connected to charge sensitive 
preamplifiers, PSC 823, produced by Canberra--France.
The preamplifiers and the cables limit the bandwidth of the system
to about 10\,MHz.
The energies and pulse forms 
were recorded
for all 19 channels
using a ``DGF Pixie-4'' data acquisition system, DAQ, 
manufactured by X-Ray Instrumentation Associates, XIA. 
The core pulse was always used to trigger.
The DAQ has 14--bit ADCs and a sampling frequency 
of 75\,MHz \cite{PIXIEmanual}.

\section{Pulse shape simulation}
\label{s:pss}

Two simulated pulses, one for the core and one for the segments, 
were used as references to determine the relative lengths and amplitudes
of measured pulses. 
They were generated using the
pulse shape simulation package \cite{pss} developed for 
true-coaxial detectors.

The pulses were simulated for a point-like energy deposit of 122\,keV
at $r=32.5$\,mm, $\phi=0^{\circ}$ and $z=0\,$mm.
This point is located on the  $\langle 110 \rangle$ axis
of the detector, 5\,mm deep inside the detector. The depth was chosen
as it is the average penetration depth of 122\,keV photons.

The parameters of the simulation were:
\begin{itemize}
  \item impurity level of $0.45 \cdot 10^{10}$\,/cm$^3$, 
        corresponding to the specification at $z=0$\,mm;
  \item electron mobility \cite{bart}, corresponding to $T=$77\,K;
  \item grid for numerical field calculations of $32(r)\times180(\phi)\times70(z)$;
  \item time, t, development in steps of 1\,ns, corresponding
        to spatial steps of about 100\,$\mu$m which is the granularity
        needed to reflect the development of charge trajectories;
  \item bandwidth of 10\,MHz and amplifier decay time of 50\,$\mu$s;
  \item no noise.
\end{itemize}

\begin{figure}[htbp]
\centering
\includegraphics[width=0.48\textwidth]{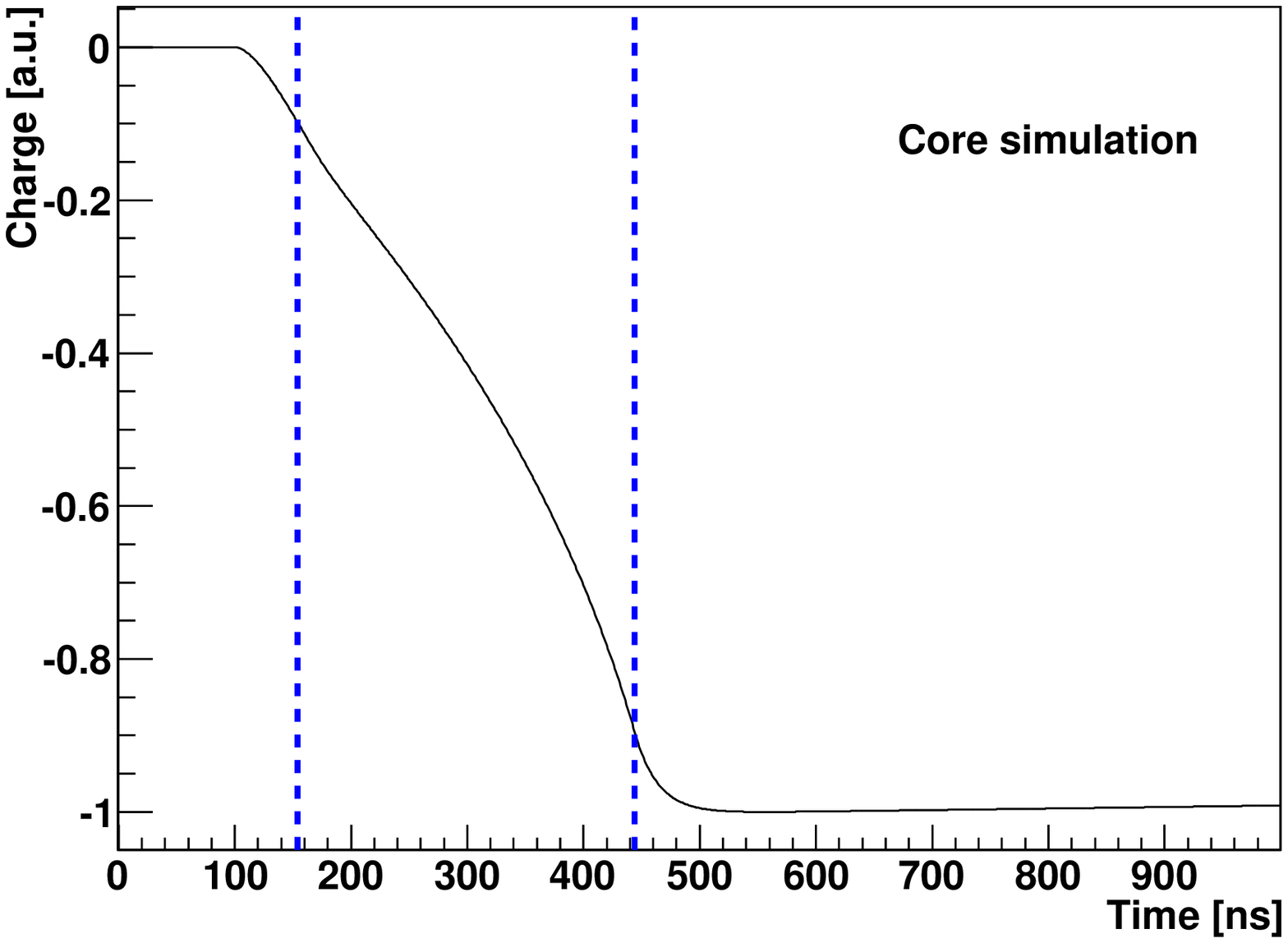}
\includegraphics[width=0.48\textwidth]{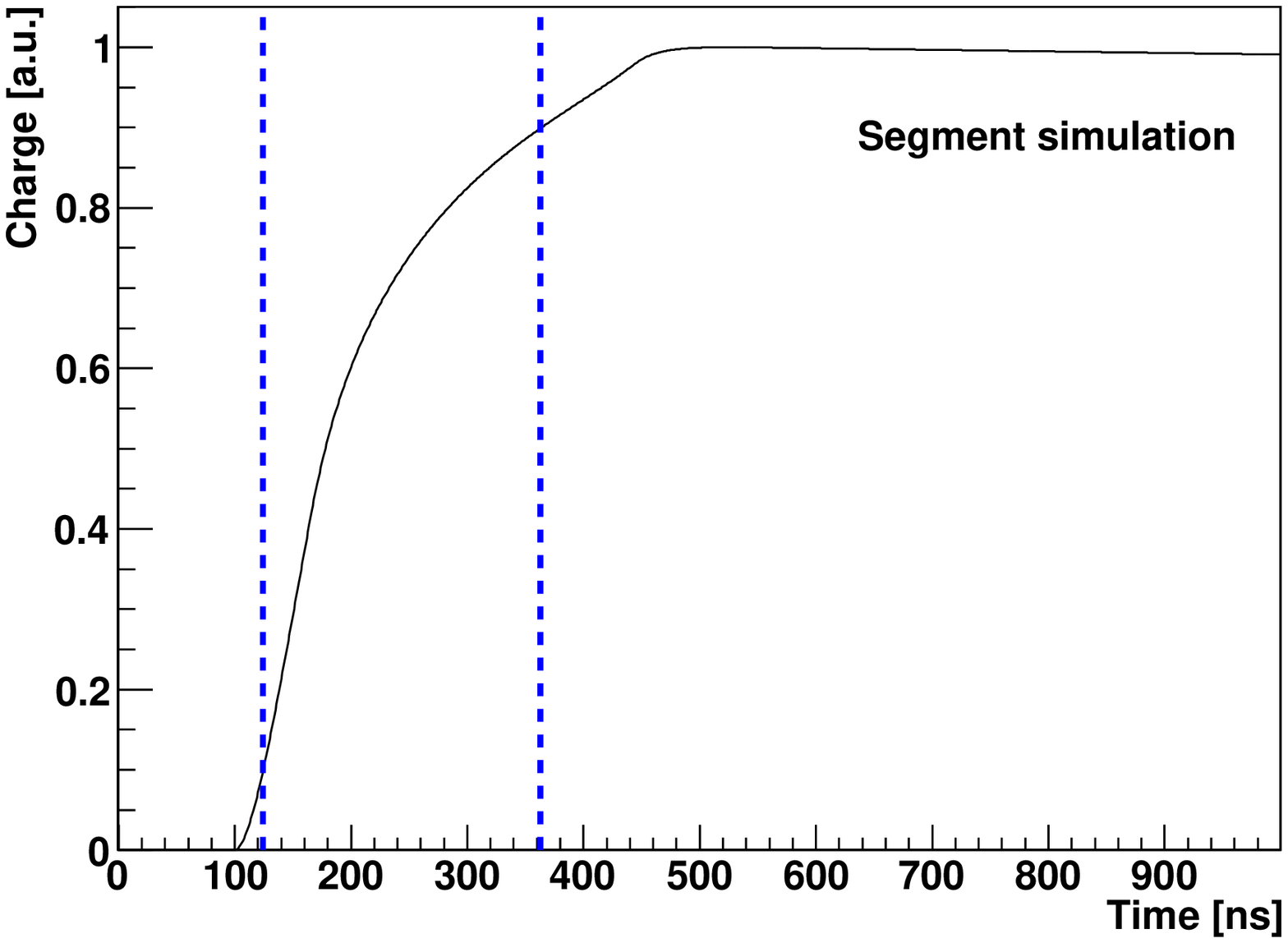}\\
\flushleft
  \vskip -0cm
  \hskip 3.5cm a)
  \hskip 6cm b)
  \vskip 0cm
\caption{Simulated reference pulses for a) core and b) segments. 
The pulses are normalized to $\pm1$. 
The times at which 10\,\% and 90\,\% of the amplitude are reached
are indicated by dashed lines.}
\label{f:simpulse} 
\end{figure}

The simulated charge pulses, $C_{\text{sim}}^{c,s}$, 
for core, $c$, and segments, $s$, 
are shown in Fig.~\ref{f:simpulse}.
The amplitude is normalized to $\pm1$, respectively.
The pulses start at t=100\,ns and have a total rise time, $t_{\text{r}}$,
of almost 400\,ns.
The times, at which the pulses
reach 10\,\% and 90\,\% of their amplitudes, are indicated. The
time interval inbetween, the so called 10--90 rise-times, 
$t_{\text{r}}^{c,s:10-90}$, differ for core and segments.
The  $t_{\text{r}}^{c,s:10-90}$ were used in all measurements   
to evaluate the length of a pulse, because the beginning and the end of
a pulse are very susceptible to noise.  

The usage of a reference pulse, 
simulated as one interaction 5\,mm inside the detector, 
instead of an average over 
pulses simulated according to beamspot and penetration depth, does not
change $t_{\text{r}}^{c:10-90}$ significantly \cite{pss}.
The effect on  $t_{\text{r}}^{s:10-90}$ 
will be subject of a separate investigation. In general, the hole drift
is less well understood than the electron drift. This is reflected in
a less accurate prediction of the detailed pulse shape for the segments.

The simulated core and segment  pulses along $\langle 110 \rangle$ at 77.4\,K
have 
$t_{\text{r}}^{c:10-90}= 290$\,ns 
and
$t_{\text{r}}^{s:10-90}= 239$\,ns.

\section{Model expectations}
\label{s:model}

The $T$ dependence of the rise time is assumed to be due to the
$T$ dependence of the mobility of the charge carriers.
The velocity, $\vec{v}_{e/h}(\vec{x})$ of the charge carrying electrons, $e$,
and holes, $h$, at each point $\vec{x}=(r,\phi,z)$ is given by 

\begin{equation} 
\label{e:para} 
\vec{v}_{e/h}(\vec{x}) = \frac{\mu_{e/h}\vec{E}(\vec{x})}
           { [1+( \frac{|\vec{E}(\vec{x})|} {E_0} )^{\beta}]^{1/\beta} }~~~, 
\end{equation}

where 
$\mu_{e/h}$ is the charge carrier mobility tensor,
$\vec{E}(\vec{x})$ is the electric field and 
$E_0$ and $\beta$ are fit parameters obtained in measurements \cite{bart}
along the crystallographic axes.
The mobility tensor depends on the relative
position of $\vec{x}$ to the crystallographic axes. Along the axes,
$\mu_{e/h}$ becomes a number.
A more detailed discussion relevant for the detector considered here
can be found elsewhere \cite{pss}. 

As the detector is true-coaxial, $\vec{E}$ is radial with
the absolute value, $E(r)$. Thus, the velocity, $v_{e/h}(r)$, becomes radial
along the crystallographic axes.
Due to the extremely low impurity density, $E(r)$ varies by
less than 50\,\% over $r$ \cite[~Fig.1]{pss}.  
Thus it is appropriate to use the approximation 

\begin{equation} 
\label{e:velocity}
v_{e/h}(r) = \mu^{\text{eff}}_{e/h} E(r) ~~,
\end{equation}

where $ \mu_{e/h}^{\text{eff}} $
are the effective mobilities of the charge carriers. 

The differential equations, 
$v_{e,h}(t)=dr_{e,h}(t)/dt$, 
for the radial drift, $r_{e,h}(t)$, of the electrons and holes
along a crystallographic axis
can be solved.
In the case considered here, the electron drift dominates and
any
$T$ dependence, $\mu_{e}^{\text{eff}}(T)$, is directly reflected in a
$T$ dependence of the rise time, $t_{\text{r}}(T)$, which becomes

\begin{equation}
\label{eq:risetime}
t_{\text{r}}(T)= \frac {C_{\text{det}}} {\mu_{e}^{\text{eff}}(T)} \,,
\end{equation}

where $C_{\text{det}} $ is a constant that was calculated from
the detector geometry and its electrical parameters after solving the
equations. For the detector under consideration it is
$C_{\text{det}} =
5.9 \cdot 10^{-3}$\,cm$^2$/V.

It is generally assumed that $\mu_{e}^{\text{eff}}(T)$
can be described as 

\begin{equation}
  \label{eq:mu}
  \mu_{e}^{\text{eff}}(T) = \mu_e^T f(T),
\end{equation}

where  $\mu_e^T$ is a constant, $f$ is some functional form
and $T$ is given in Kelvin.
Models based on the scattering of 
electrons and holes off the phonons
in the lattice \cite{T32} suggest a form of
$f(T)=T^{-3/2}$.
This is qualitatively supported by some  
measurements at higher impurity levels \cite[p.~108]{Semicond}.
Other measurements, also at higher impurity levels, show
a behavior of $f(T)=T^{-1.6}$ or $f(T)=T^{-1.66}$ \cite{T16}.

The values of the electron mobility used 
in the simulation
for the $\langle 110 \rangle$ axis at 77\,K 
can be translated into a value for
$\mu_e^{\text{eff}}(77$\,K) of $1.4\cdot10^4$\,cm$^2$/(Vs).
For $f(T)=(T/\text{K})^{-3/2}$, the constant $\mu_e^T$ becomes 
$\mu_e^T = 9.4 \cdot 10^{6}$\,cm$^2$ /(Vs).
This provides an absolute
model prediction to be compared to data.
As it is not expected that the shapes of the pulses 
for core and segment vary significantly within the T range,
it also provides a prediction for $t_{\text{r}}^{c,s:10-90}$.

\section{Event selection}
\label{s:evsel}

\begin{figure}[htbp]
\centering 
\includegraphics[width=0.8\textwidth]{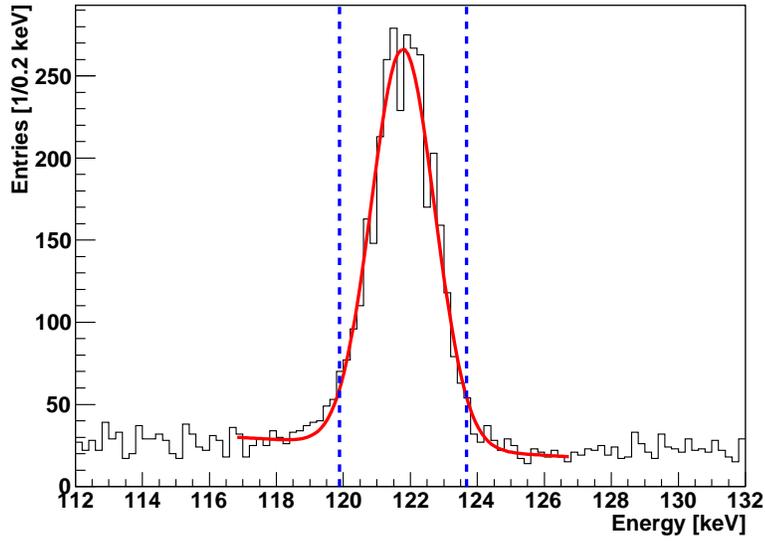}
\caption{Energy spectrum around the 122\,keV peak
as measured by the segment on 
the $\langle 110 \rangle$ axis. 
Also shown is a fit with a Gauss function 
plus a first-order polynomial. 
The $2\sigma$ limits are shown as dashed lines.}
\label{f:spect} 
\end{figure}

Events induced by 122\,keV photons were selected using the 
energies measured in the individual segments.
Single segment events were selected by requiring energy deposits of less
than 20\,keV in all other segments.
Figure~\ref{f:spect} shows the spectrum 
around 122\,keV of the segment containing the
$\langle 110 \rangle$ axis 
for single segment events from DS3 
at $T_{\text{mon}}$=104\,K.
The width of the peak was determined by a fit of a Gauss function
and events within $\pm 2\sigma$ were selected.
Typical signal to background ratios were 3:1 to 5:1 for DS2 and DS3
and 1:1 to 2:1 for DS1.

\section{Rise time determination}
\label{s:ple}

The rise time of a measured pulse was not extracted directly because of the
noise level observed during the measurements.
Instead, a fitting procedure was used, where a simulated pulse was fitted
to the measured pulse as a reference. 
This method was chosen, because it uses the full information
of the measurement
and is robust against noise which averages out.

\begin{figure}[htbp]
\centering 
\includegraphics[width=0.8\textwidth]{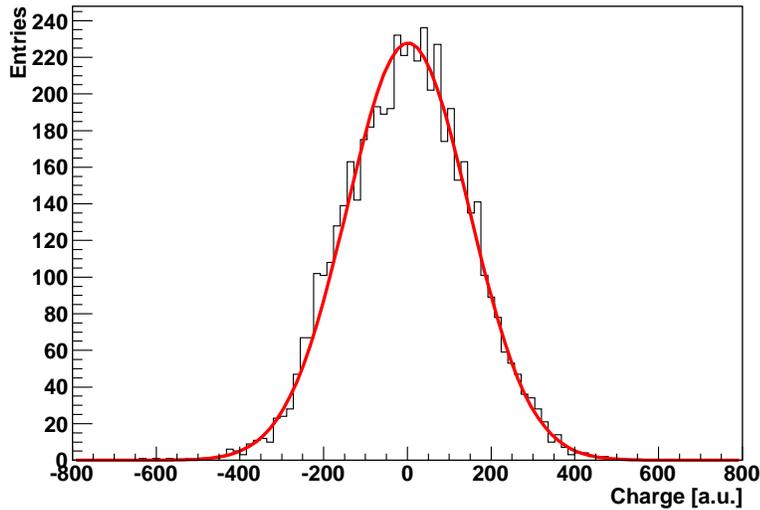}
\caption{Typical noise distribution in arbitrary units. 
         Also shown is a Gaussian fit.}
\label{f:noise} 
\end{figure}

Before the fit, the baseline was subtracted from each pulse.
It was calculated as the average of the values recorded 
for the first  15 bins, 200\,ns, of the pulse record. This was possible,
because the DAQ was adjusted to record pulse information
starting 200\,ns before the trigger time.

The noise level for each 15--minute interval
was determined using the first 200\,ns
of several hundred baseline subtracted pulses. 
Figure~\ref{f:noise} shows a typical noise distribution. The unit of the
charge is arbitrary. The noise is clearly Gaussian. 
Its level was typically about 3\,\% of the pulse amplitude.
Each point of a measured pulse was assigned an uncertainty corresponding
to the noise level of the dataset.

The simulated pulses, $C_{\text{sim}}^{c,s}$, were fitted separately 
to the measured 
core and segment pulses:

\begin{equation}
\label{eq:simpulse_param}
   C_{\text{meas}}^{c,s}(t)=A^{c,s}\cdot C_{\text{sim}}^{c,s}(t / t_{\text{scale}}^{c,s} + t_{\text{offset}}^{c,s}),
\end{equation}

where the fitted parameters are the relative amplitudes,
$A^{c,s}$, 
the time offsets of the measured pulses,
$t_{\text{offset}}^{c,s}$, 
and the time scaling factors
$t_{\text{scale}}^{c,s}$. 
The important parameters are $t_{\text{scale}}^{c,s}$ which  
were converted into measured  values of $t_{\text{r}}^{c,s:10-90}$.

\begin{figure}[htbp]
\centering 
\includegraphics[width=0.7\textwidth]{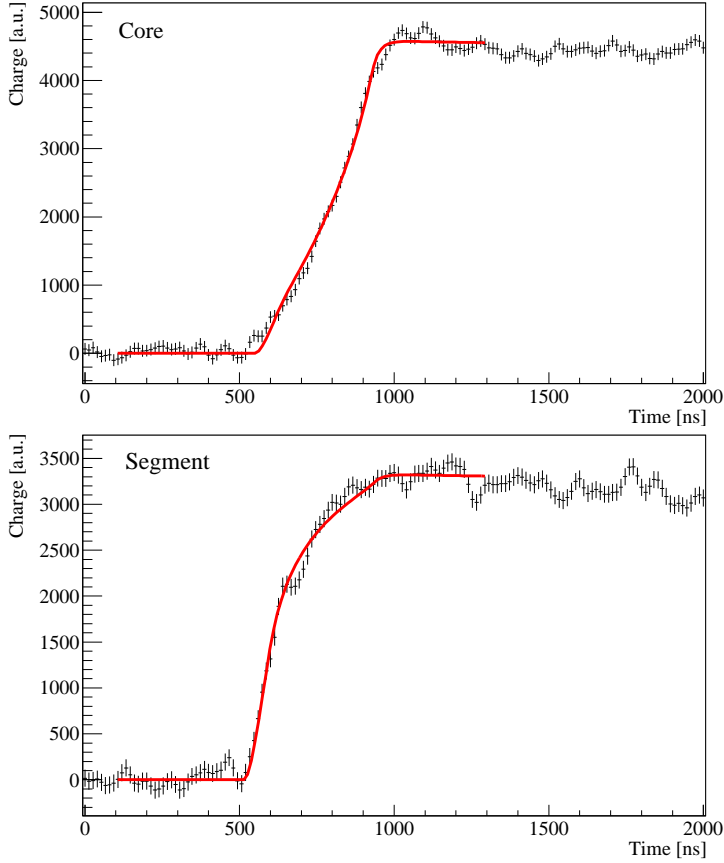}
\caption{Pulses in  core and  segment of a typical event
         from DS3 at $T_{\text{mon}}=100$\,K. 
         The fits with the simulated pulses are also shown.
         The  fits  are classified as good with 
         $\chi^2/\mbox{ndf}=1.06(1.05)$ for the core(segment).} 
\label{f:fitgood} 
\end{figure}

\begin{figure}[htbp]
\centering 
\includegraphics[width=0.7\textwidth]{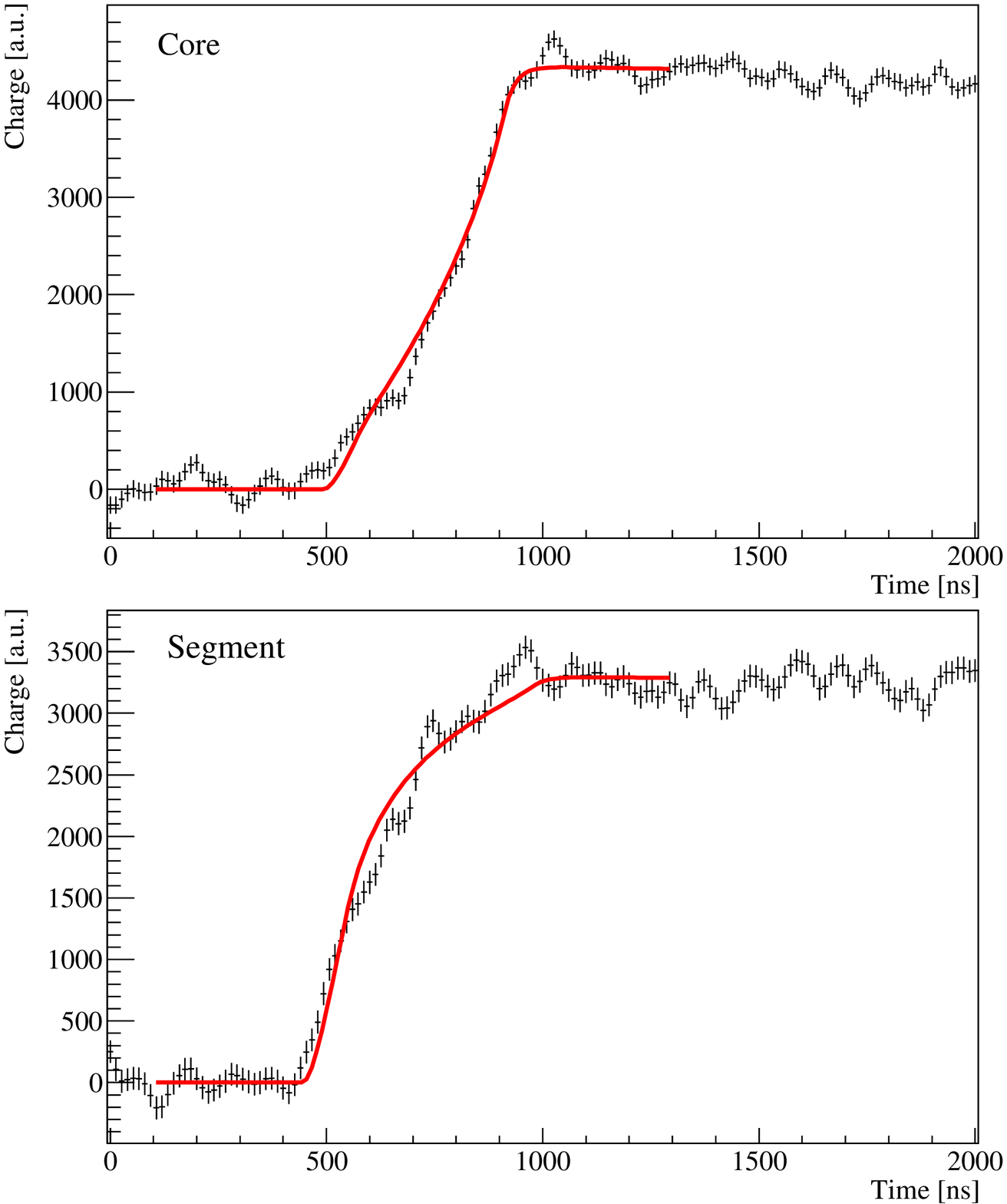}
\caption{Pulses in  core and segment of a typical event
         from DS3 at $T_{\text{mon}}=100$\,K.  
         The fits with the simulated pulses are also shown.
         The  fits are classified as bad with 
         $\chi^2/\mbox{ndf}=1.87(2.29)$ for the core(segment).}
\label{f:fitbad} 
\end{figure}

Only good fits were considered.
A fit was qualified as good, if 
$\chi^2/\mbox{ndf}\leq \chi^2_{c,s}$.
The values for $\chi^2_{c,s}$
were chosen to reject 50\,\% of the pulses corresponding
to the worst background levels observed. They were adjusted for each
dataset and where 1.1(2.0) for DS1, 1.2(1.7) for DS2 and 
1.7(1.9) for DS3 for the core (segment).
Figures~\ref{f:fitgood} and~\ref{f:fitbad} show examples of two events
with good and  bad fits for both core and segment pulses.

\begin{figure}[htbp]
\centering 
\includegraphics[width=0.6\textwidth]{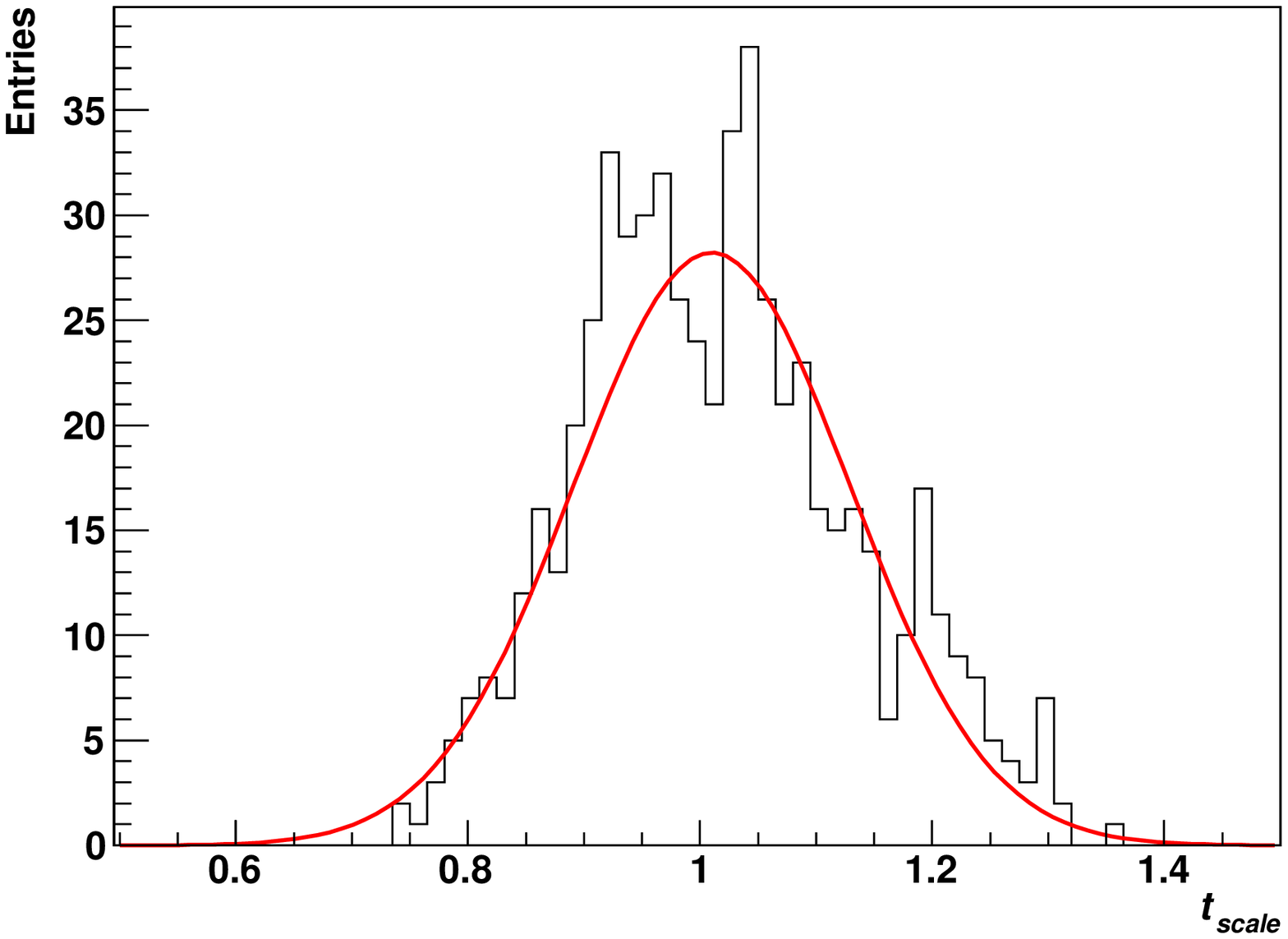}
\caption{Distribution of the scaling parameter $t^c_{\text{scale}}$
         for measured core pulses along the $\langle 110 \rangle$ at  77.4\,K.
         Also shown is a Gaussian fit.}
\label{f:tscale} 
\end{figure}

A bad fit could indicate that the interaction point was not close to the
outer mantle. This can happen when a 122\,keV photon penetrates deeper
than in average or, if the event results from an interaction associated
to the Compton background. The background also contains multiple
interactions.

The distributions of the fit parameters $t_{\text{scale}}^{c,s}$ 
for all good fits were fitted with
Gauss functions and the resulting means, $t_{\text{scale-mean}}^{c,s}$,  
are converted to measured $t_{\text{r}}^{c,s:10-90}$ values
for a given position and temperature.

Figure~\ref{f:tscale} shows the $t^c_{\text{scale}}$ distribution for
the core pulses along the $\langle 110 \rangle$ axis at  77.4\,K.
The resulting $t_{\text{scale-mean}}^{c}$ is very close to 1 and results
in a measured pulse length of $t_{\text{r}}^{c:10-90}=287\pm$2\,ns. 
The uncertainty is statistical only. 

The measured value for the segment is
$t_{\text{r}}^{s:10-90}=305\pm$ 2\,ns.
The agreement between simulation and measurement
is excellent 
for the core 
while it is quite poor
for the segments. 
The discrepancy might be explained by a combination of the poor 
understanding of the hole mobility and net charge carrier density 
variations close to the core.
The hole mobility determines the time when the 10\,\% level is reached. 
The 90\,\% level could be influenced by a changing drift velocity of the
electrons close to the core due to changing impurity levels.
The segment pulses would be affected more than the core pulses due
to the different strengths of weighting fields.
The relative discrepancy should, however, not change within the temperature
range considered.

\section{\boldmath$T$ dependence}
\label{s:tdep}

The values of $t_{\text{r}}^{c,s:10-90}$
as determined according to the procedure described in the
previous section were used to study the temperature dependence
of the rise time.
The analysis was performed separately for the two crystallographic
axes $\langle 110 \rangle$ and $\langle 100 \rangle$
using the datasets A1 and A2 from Table~\ref{t:DS}.
The drift along $\langle 100 \rangle$ is 
known to be faster at 77\,K \cite{bart}. The $T$ dependence is $a~priori$ assumed
to be the same.

As explained in section~\ref{s:exp}, $T$ is only
known up to a measurement dependent shift $\Delta T$
for DS2 and DS3. When a functional form $f$, 
as introduced in Equation~\ref{eq:mu},
was fitted to rise-time data including DS2 or DS3, free parameters 
$\Delta T^{2,3}$ were
introduced to the fit, respectively.

\begin{figure}[htbp]
\centering 
\includegraphics[width=0.9\textwidth]{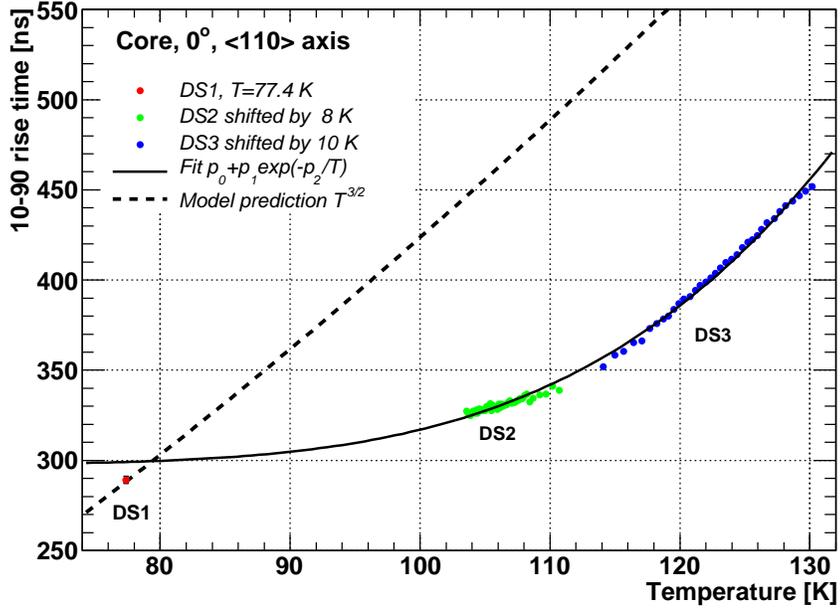}
\caption{Dependence of $t^{c:10-90}_{\text{r}}$ 
         on $T$ along the $\langle 110 \rangle$ axis.
         Also shown are the model prediction described in 
         section\,\ref{s:model} and a fit with a Boltzmann-like
         $ansatz$.
         Statistical uncertainties are smaller than the symbol size.
}
\label{f:RT110c} 
\end{figure}

Figure~\ref{f:RT110c} shows  the development of
$t_{\text{r}}^{c:10-90}$  for the $\langle 110 \rangle$ 
axis together with the model prediction 
from section~\ref{s:model},

\begin{equation}
\label{eq:modela}
   t_{\text{r}}(T)= \frac {C_{\text{det}}} {\mu_{e}^T} T^{3/2} ~~,
\end{equation}
 
and a fit with a simple Boltzmann-like $ansatz$
 
\begin{equation}
\label{eq:ansatz}
   t_{\text{r}}(T)= p_0+p_1 e^{-\,p_{2}/T}~~,
\end{equation}
where $p_0,~p_1,~p_2$ are free parameters.

\begin{figure}[htbp]
\centering 
\includegraphics[width=0.9\textwidth]{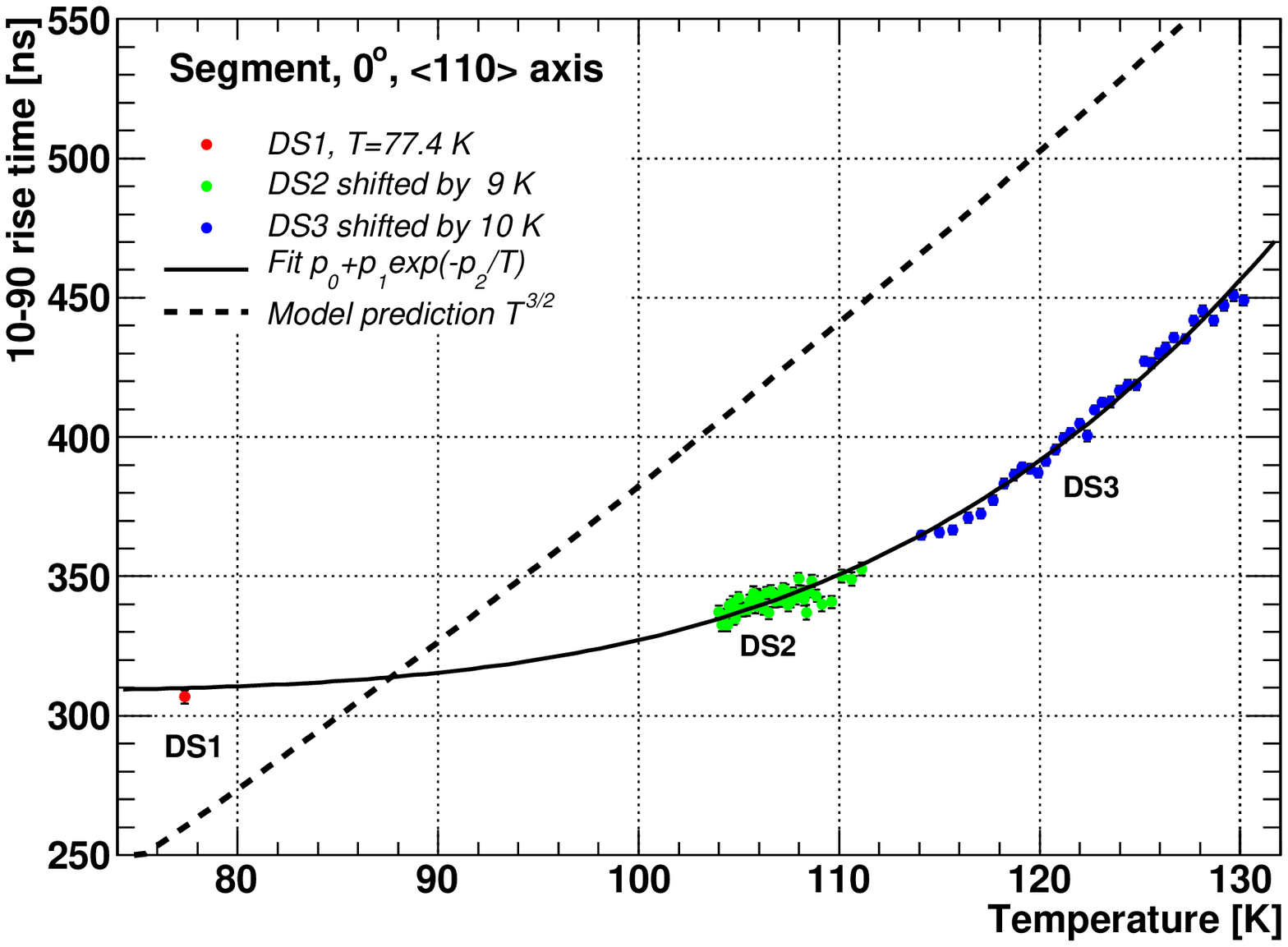}
\caption{Dependence of $t^{s:10-90}_{\text{r}}$ 
         on $T$ along the $\langle 110 \rangle$ axis.
         Also shown are the model prediction described in 
         section\,\ref{s:model} and a fit with a Boltzmann-like
         $ansatz$.
         Statistical uncertainties are shown, but are mostly
         smaller than the symbol size.
}
\label{f:RT110s} 
\end{figure}

The model prediction does not describe the data at all.
It correctly predicts the value of $t_{\text{r}}^{c:10-90}$
at 77.4\,K, but
it is clear that the predicted $T$ dependence is not observed in the
data.
The decription cannot be improved by modifying the exponent.
The more general
$ansatz$ $t_{\text{r}}(T)=p_0+p_1 T^{p_2}$ 
can also not describe the data. The description is bad
in the $T$ range of DS2 and DS3 and completely misses the DS1 point.
However, the Boltzmann-like $ansatz$ describes the data quite well.
This is also true for the segment pulses as shown in Fig.~\ref{f:RT110s}.
For the segment pulses, the model prediction 
from section~\ref{s:model} neither describes the $T$ dependence nor
the point at 77\,K.
The hole drift is more important in this case and as mentioned in
section~\ref{s:pss} the parameters for the hole drift are less well
measured.

\begin{figure}[htbp]
\centering 
\includegraphics[width=0.9\textwidth]{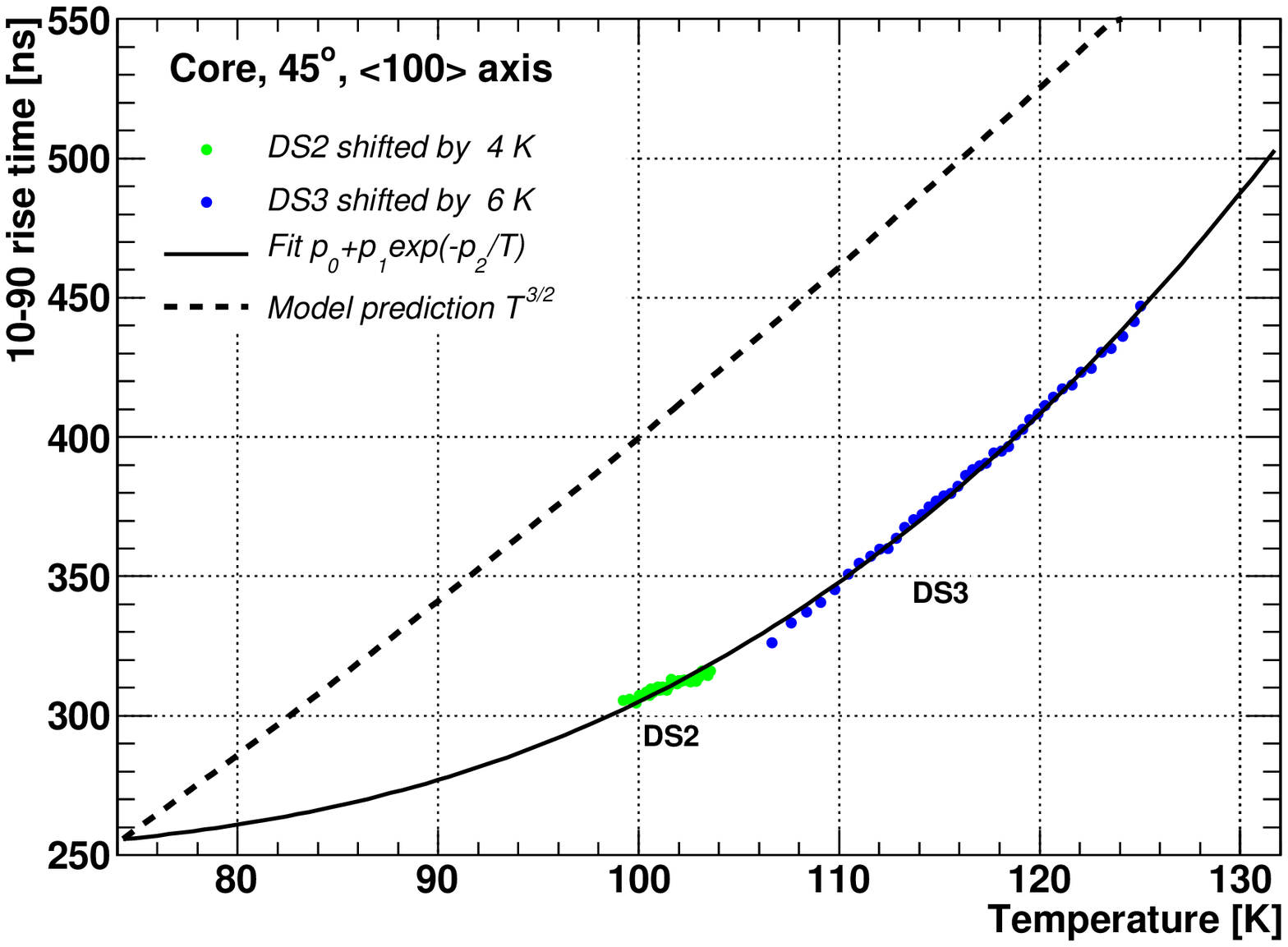}
\caption{Dependence of $t^{c:10-90}_{\text{r}}$ 
         on $T$ along the $\langle 100 \rangle$ axis.
         Also shown are the model prediction described in 
         section\,\ref{s:model} and a fit with a Boltzmann-like
         $ansatz$.
         Statistical uncertainties are smaller than the symbol size.
}
\label{f:RT100c} 
\end{figure} 

\begin{figure}[htbp]
\centering 
\includegraphics[width=0.9\textwidth]{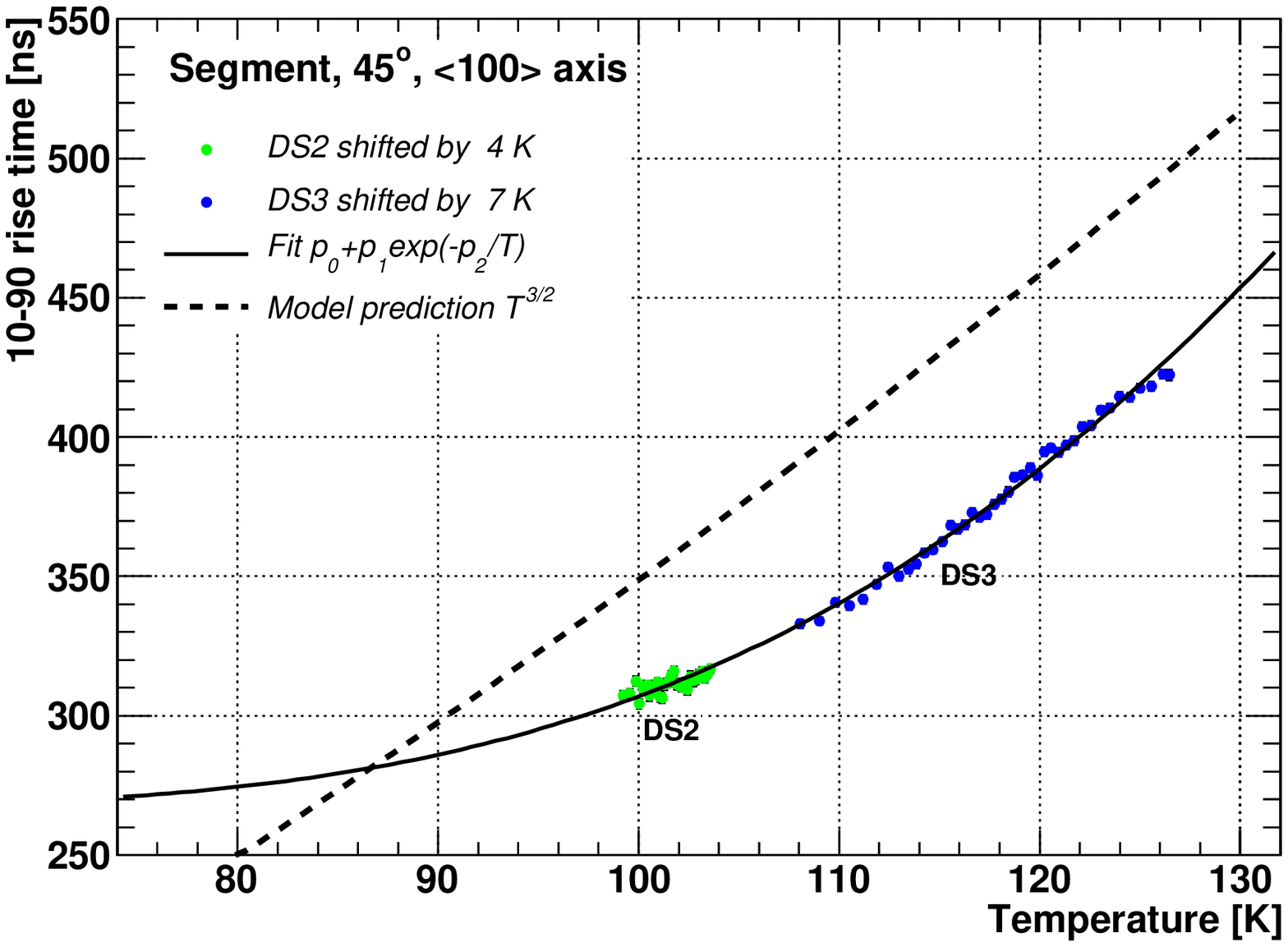}
\caption{Dependence of $t^{s:10-90}_{\text{r}}$ 
         on $T$ along the $\langle 100 \rangle$ axis.
         Also shown are the model prediction described in 
         section\,\ref{s:model} and a fit with a Boltzmann-like
         $ansatz$.
         Statistical uncertainties are shown, but are mostly
         smaller than the symbol size.
}
\label{f:RT100s} 
\end{figure}

The situation for the $\langle 100 \rangle$ axis
is shown in Figs.~\ref{f:RT100c} and~\ref{f:RT100s}.
No data were available at 77\,K. 
However, it is clear
that a power law cannot describe the data 
for either core or segment, while a Boltzmann-like $ansatz$ 
can be fitted quite well.

The success of the 
Boltzmann-like $ansatz$ could be connected to a change in the 
conductivity of the germanium crystal \cite{IKZ}
in the $T$ range considered. 
This  might have affected the 
electric field strongly, 
because the detector was operated relatively
closely to the full depletion voltage.

Table~\ref{t:p2} lists the values obtained 
in $\chi^2$--fits 
for the parameters 
$p_1$ and $p_2$ from Equation~\ref{eq:ansatz}
which are strongly correlated.
The values for $p_2$ are also shown as energies, 
$E$, with $p_2= E/(2k)$, where
$k$ is the Boltzmann constant. The uncertainties given are 
statistical only. The $\chi^2$/ndf is typically around 5.
 
\begin{table}[!h]
  \caption{Fit parameters determined for a Boltzmann-like $ansatz$
           using the datasets A1, A2, C1 and C2}
  \vskip 0.2cm
  \centering
    \begin{tabular}{ l | l| c | r | l | r@{}@{}l}
      \hline
      \hline
         & orientation & $c,s$ & \multicolumn{1}{|c|}{$p_1$\,
[$\cdot$10$^4$]}&\multicolumn{1}{|c|}{$p_2$} & \multicolumn{2}{|c|}{$E$
[meV]} \\
      \hline
      \hline
      A1& $\langle 110 \rangle$ & $c$ & 17.7 $\pm$ 1.3  &  913 $\pm$
10   & 158\, & $\pm$\,2 \\
      \hline
      A1& $\langle 110 \rangle$ & $s$ & 15.0 $\pm$ 2.3  &  900 $\pm$
20   & 155\, & $\pm$\,4 \\
      \hline
      A2& $\langle 100 \rangle$ & $c$ & 3.0 $\pm$ 0.2 &  630 $\pm$
~\,8 & 109\, &$\pm$\,2 \\
      \hline
      A2& $\langle 100 \rangle$ & $s$ & 3.1 $\pm$ 0.4   &  666 $\pm$
18   & 115\, &$\pm$\,3 \\
      \hline
      \hline
C1& $\langle 110 \rangle$ +5\degree & $c$ & 3.5 $\pm$ 0.3 &  649 $\pm$
12   & 112\, &$\pm$\,2 \\
      \hline
C1& $\langle 110 \rangle$ +5\degree & $s$ & 5.6 $\pm$ 0.7 &  725 $\pm$
18   & 125\, &$\pm$\,3 \\
      \hline
C2& $\langle 110 \rangle$ -15\degree& $c$ & 5.5 $\pm$ 0.5 &  705 $\pm$
13   & 122\, &$\pm$\,2 \\
      \hline
C2& $\langle 110 \rangle$ -15\degree& $s$ & 5.5 $\pm$ 0.9 &  732 $\pm$
22   & 126\, &$\pm$\,4\\
      \hline

    \end{tabular}
  \label{t:p2}
\end{table}

The core and segment parameters agree 
within statistical uncertainties 
for the datasets A1 and A2, respectively.
However, the parameters for the two axes seem to be 
significantly different.
A cross-check for the $\langle 110 \rangle$ axis, using 
dataset C1 from Table~\ref{t:DS},
did, however, not confirm this.
The parameters should be the same, but they agree better
with the parameters of the fit to A2.
The inclusion of DS2 is what pushes the values
of $p_1$ and $p_2$ up in the fit to A1.
As the absolute temperatures for DS2 and DS3 are not known, 
but have to be fitted,
we used this cross-check to estimate the systematic uncertainty 
on $E$ to be $\approx$\,40\,meV.
The systematic uncertainty could also be connected to 
the misalignment present for DS3,
even though such a strong deviation is not expected for a 5\degree
offset.
Another cross-check was performed using the dataset C2.
At that position, the drift is not simply radial any more
due to the distance from the crystallographic axes \cite[~Fig.3]{pss}, 
but the data could still be described by the Boltzmann-like $ansatz$.
The fit results for C2 and C1 are compatible.

The fits were further analysed using the program package BAT \cite{BAT}.
The values listed in Table~\ref{t:p2} were confirmed by BAT as the global
mode. A detailed study of the probability densities revealed that the
inclusion of DS2 causes elongated two-dimensional probablity densities
with local minima, also pointing to systematic uncertainties of up to
$\approx$\,40\,meV for the fits to A1 and A2.
The probability densities for the fits to C1 and C2 are much more regular,
pointing at systematic uncertainties of less than $\approx$\,20\,meV.

The drift along the two crystallographic axes were 
expected to have the same $T$ dependence.
The data, independent of fitting procedures,
 seem to indicate that the $T$ dependence of the
``slow axis'', $\langle 110 \rangle$, is less pronounced.
At some $T$, the rise time along the ``fast axis'', $\langle 100 \rangle$,
seems to become as large as along the ``slow axis''. 
A definite conclusion would need better data.
More measurements at more angles and with an improved temperature monitoring
are planned. 

It should be noted that
while the rise times changed with temperature, the amplitudes remained stable,
even to the highest temperature of about 130\,K. This means that
the detector could be operated stably at temperatures well above 100\,K.

\section{Summary}

The temperature dependence of the rise time of pulses was studied between
77\,K and 130\,K. 
The data cannot be described by the expected  $T^{-3/2}$ dependence
of the charge carrier mobilities.
Both core and segment pulses display a Boltzmann-like
temperature dependence of the rise time.

The significant temperature dependence of the rise time 
should be taken into account
when pulses are simulated for specific experimental setups
or used in pulse shape analyses.
It should be noted that the detector used for the investigation could
be stably operated up to temperatures of 130\,K without an observable
change in the energy scale.



\end{document}